\documentclass[aps,pre,twocolumn,showpacs]{revtex4}
\usepackage{amssymb}
\usepackage{graphicx}
\usepackage{amsmath}
\begin{document}
\title{L\'{e}vy  flights in inhomogeneous environments}
\author{Piotr Garbaczewski and Vladimir Stephanovich}
\affiliation{Opole University, Institute of Physics, 45-052 Opole, Poland}
\begin{abstract}
We study the long time asymptotics of probability density functions
(pdfs) of  L\'{e}vy flights in different confining potentials. For
that we use two models: Langevin - driven and (L\'{e}vy -
Schr\"odinger) semigroup - driven dynamics. It turns out that the
semigroup modeling  provides much stronger confining properties than
the standard Langevin one. Since contractive semigroups set a link
between L\'{e}vy flights and fractional  (pseudo-differential)
Hamiltonian systems, we can use the latter  to  control the long -
time asymptotics of the  pertinent  pdfs. To do so, we need to
impose suitable restrictions upon the Hamiltonian and its potential.
That provides verifiable  criteria for an invariant pdf to be
actually an asymptotic pdf of the semigroup-driven  jump-type
process.   For computational  and visualization  purposes  our
observations are  exemplified for the  Cauchy driver and its
response to  external polynomial  potentials (referring to L\'{e}vy
oscillators), with respect to both dynamical mechanisms.
 \end{abstract}
  \maketitle
\section{Introduction}

In recent years there has been growing interest in random walks,  extending from  various fields of physics to  chemistry, biology and
financial mathematics. The classical concept of Brownian motion has become paradigmatic in the whole theory of stochastic
processes, see e.g. Ref. \cite{lev1}. The probability density function (pdf) of a homogeneous Brownian motion solves
 a Fokker-Planck equation and has an important intrinsic property: the diffusing particle pdf, that is  initially concentrated
 at a point, with the flow of time takes the Gaussian form, whose width grows in time as $t^{1/2}$.
 This kind of diffusion processes was called the normal diffusion.
 We leave aside  a broad  field of anomalous diffusions, where $<X^2(t)>  \sim  t^{\alpha }$ with $0<\alpha <2$,
 and focus on non-Gaussian, jump-type stochastic processes, whose pdfs belong to a class of  L\'{e}vy - stable distributions.
 They have long tails and    their  second (and higher) moments are nonexistent. The existence of the first moment is granted
 only for a suitable subclass  \cite{lev2}.

In the  standard theory of diffusion-type processes, a celebrated
method to solve transport problems with respect to the  pdf and
specifically transition probability density,  is to transform the
involved Fokker-Planck equation into  its  parabolic (Hermitian)
counterpart, \cite{risken}. Right at this point a dynamical
(Schr\"odinger) semigroup notion naturally appears and  its  obvious
link with a Hamiltonian dynamical system. Indeed,  one encounters
here  the  Schr\"{o}dinger-type equation (there is no imaginary unit
before time derivative and physical dimensions of the Hamiltonian
are re-scaled accordingly), often called the generalized diffusion
equation, $\partial _t\Psi = \hat{H}\Psi $.  Here,  $\hat{H}$ is
interpreted as  a Hamiltonian operator and  $-\hat{H}$  stands for a
generator of the dynamical semigroup  $\exp(-t\hat{H})$.

In case of Brownian motion and normal diffusion process, the  Fokker
- Planck (Langevin - driven) and semigroup dynamics refer to the
same random  process. A usefulness of the the semigroup picture lies
in the possibility of eigenfunction expansions of $\hat{H}$ which
allows to deduce explicit formulas for transition  (semigroup)
kernels and transition pdfs, \cite{risken}.  As a byproduct of the
above procedure one solves the eigenvalue problem for the
Hamiltonian operator and identifies its ground state   as a square
root of an invariant pdf of the stochastic diffusion process. The
latter pdf is approached by the process at large times.

The situation is somewhat different in the case of non-Gaussian
jump-type processes. Namely, the (L\'{e}vy - Schr\"{o}dinger)
semigroup  dynamics differs from the Langevin - driven fractional
Fokker-Planck  evolution, see e.g. \cite{stef,sokolov,geisel}.
Nonetheless a common  asymptotic  invariant pdf may be attributed to
both dynamical scenarios  \cite{stef}.  Since Langevin-driven
dynamics does not admit invariant pdfs  in the Gibbs form (e.g.  the
force potential  does not appear in the exponential form of $\rho
_*$), the whole class of
 pdfs, originally  employed in the study of topologically induced super-diffusions \cite{sokolov,geisel},
 might seem to be  excluded by the formalism of Ref. \cite{stef}.  This is not the case.

In the present paper we focus on extending the range of validity of the  reverse engineering (targeted stochasticity)
problem of Ref. ~\cite{klafter} to so-called topological L\'{e}vy processes (topologically induced super-diffusions),
occurring in systems with topological complexity like folded polymers and complex networks. To be more specific, the
 original  reverse engineering problem formulates as follows: given an invariant pdf $\rho _*(x)$, design a stochastic
 Langevin-driven jump-type process for which the preselected density may be an  {\it  asymptotic}  target.
 The basic reconstruction goal  is to deduce the drift function of the process.

In the previous paper, \cite{stef}, we have  recast  the reverse
engineering  problem so that  the original  task was supplemented by
one more   reconstruction step. Namely, we have addressed  the
existence issue  of
 the  L\'{e}vy - Schr\"odinger semigroup potential
${\cal{V}}(x)=-\lambda \, (|\Delta|^{\mu/2}\rho _*^{1/2})(x)\, /\rho _*^{1/2}(x)$
(see below for detailed explanation),  given the \it  very  same  \rm (as for the Langevin process) invariant pdf $\rho _*(x)$,
that is non-Gibbsian by construction.

Presently,  we relax the previous (common pdf) constraint and address  a fully fledged reconstruction  problem
for the semigroup dynamics: given an invariant pdf, identify   the semigroup-driven  L\'{e}vy process for which the  prescribed
 invariant pdf $\rho _*$  may stand for an asymptotic one. Then, Gibbsian densities appear to be admissible and a class of jump-type processes,
 that  respond  to environmental inhomogeneities,  becomes largely extended. The corresponding jump-type processes are identified by us as
 topological processes  due to their  links with topologically-induced super-diffusions, \cite{sokolov,geisel}.

  At this point lest us stress that it has never been settled that the invariant pdfs of a topological process
   actually are  the  {\it  proper  asymptotic} ones, e.g. can be reached  in the large time asymptotics
    irrespective from a particular choice of initial data.  To this end one must resort to the contractive semigroup notion.
     One of the aims of our present discussion is to carefully check this point.

The expected asymptotic behavior may not persist for an unrestricted initial  pdf   choice.  The  signature, of whether
 such behavior is allowed or prohibited by the semigroup dynamics,   is encoded in the functional form of a semigroup potential ${\cal V}(x)$.
  The latter  needs to be reconstructed  from the  target   pdf $\rho _*(x)$,  by means of  the  above generalized reverse
  engineering problem and respect  a number of restrictions.
Minimal requirements upon the associated  pseudo-differential Hamiltonian and its  potential  ${\cal V}(x)$ were
 set in Ref.  \cite{olk}, where an explicit construction of Cauchy semigroups has been carried out. First explicit
  examples of appropriate potentials  were found in \cite{stef}.  It is the contractive semigroup dynamics that guarantees a proper
  asymptotic behavior of inferred time-dependent pdfs, c.f. also \cite{mackey} for a more advanced exposition of  that issue.

\section{L\'{e}vy  semigroups in a random motion}
\subsection{Brownian pre-requisites}

If we have a one-dimensional Smoluchowski diffusion process \cite{risken} with an initial pdf
$\rho_0(x)$, then its time evolution is determined by the Fokker-Planck equation $\partial _t\rho = D\Delta \rho - \nabla \, ( b \cdot \rho )$  where $D$ is a diffusion coefficient and the time - independent  drift $b(x)=f(x)/m\beta = - (1/m\beta )  \nabla V(x)$  is  induced by an external (conservative,  Newtonian) force field  $f(x)= - \nabla V(x)$.
We adopt a standard form $D=k_BT/m\beta $ of the diffusion coefficient, where $m$ and $\beta$ are, respectively, a mass and a reciprocal relaxation time of a particle.

Following a standard procedure \cite{risken}  we may identically rewrite the Fokker-Planck equation in terms of  an associated Hermitian (Schr\"{o}dinger-type)  problem by means of a redefinition
 \begin{equation} \label{un1}
 \rho (x,t) = \Psi (x,t) \rho _*^{1/2}(x)
 \end{equation}
that takes the Fokker-Plack equation into a parabolic one, often called a generalized diffusion equation:
\begin{equation} \label{un2}
\partial _t\Psi= D \Delta \Psi - {\cal{V}} \Psi \,
\end{equation}
for a positive function $\Psi (x,t)$. The  auxiliary  potential ${\cal{V}}$  derives  from a compatibility condition
${\cal{V}}(x) =  D \Delta \rho _*^{1/2}/\rho _*^{1/2}$,   whose equivalent  form reads
$ {\cal{V}}(x) = (1/2)[b^2/(2D) + \nabla b]$.

If the  ($1/2mD$ rescaled) Schr\"{o}dinger-type  Hamiltonian $\hat{H} = -D\Delta + {\cal{V}}$
is a self-adjoint operator in a suitable Hilbert space, then one arrives at a  dynamical semigroup
$\exp(-t\hat{H})$. We note here, that the Schr\"{o}dinger semigroup (parabolic) reformulation of the Fokker - Planck
equation  is merely another  mathematical "face" of the  diffusion process, the operator $\hat{H}$ is just one more form of
the Fokker-Planck operator \cite{sokolov}. The semigroup is contractive, hence
asymptotically $\Psi (x,t)|_{t \to \infty} \to \rho_*^{1/2}(x)$. Accordingly,  $\rho (x,t)|_{t \to \infty}\to \rho _*(x)$.

We note that for ${\cal{V}}={\cal{V}}(x)$ bounded from below,  the  integral kernel $k(y,s,x,t)=\{ \exp[-(t-s)\hat{H}] \} (y,x)$, $s<t$,
of the dynamical semigroup $\exp(-t\hat{H})$,  is  positive \cite{olk0}. The semigroup dynamics reads:
$\Psi  (x,t) = \int \Psi (y,s)\, k(y,s,x,t)\,  dy$ so that for all $0\leq s<t$
\begin{equation} \label{un3}
\rho (x,t) =  \rho _*^{1/2}  (x)\Psi (x,t) = \int p(y,s,x,t) \rho (y,s)
dy,
\end{equation}
where
\begin{equation}\label{un4}
p(y,s,x,t) = k(y,s,x,t)\frac{\rho _*^{1/2}(x)}{ \rho _*^{1/2}(y)}
\end{equation}
is the transition probability density of the pertinent Markov process. Its unique  asymptotic invariant  pdf is  $\rho _*(x)$.

For the  familiar Ornstein-Uhlenbeck version of the Smoluchowski process, the drift  is a linear function of $x$, e.g.
$b(x)=- \gamma x$,  $\gamma \equiv \kappa /m\beta $, $\kappa >0$.
The Fokker - Planck equation $\partial _t \rho = D\Delta \rho + \gamma \nabla (x\, \rho )$
supports an   invariant density
\[
\rho _*(x)= \left({\frac{\gamma }{2\pi D}}\right)^{1/2}\, \exp
\left( - {\frac{\gamma }{2D}} x^2\right) = \exp\left(\frac{F_* -
V(x)}{k_BT}\right)\, ,
\]
where $V(x)  = \kappa {\frac{x^2}2}$   and $F_* =  - k_BT \ln (2\pi
k_BT/\kappa)^{1/2}$. The associated   generalized heat equation
involves $\hat{H} = -D\Delta + {\cal{V}}$ with ${\cal{V}}(x) =
{\frac{\gamma ^2x^2}{4D}}  - {\frac{\gamma }2}$ which is is a
typical confining  potential. Accordingly, $\hat{H} \rho
_*^{1/2}=0$. The lowest eigenvalue  $0$ of this positively defined
operator identifies $\rho _*^{1/2}$ as its ground state function.
Eq. \eqref{un4} rewrites as
\begin{equation} \label{un4a}
p(y,s,x,t)=  k(y,s,x,t) \exp  [V(y)- V(x)]/2k_BT\, .
\end{equation}

\subsection{Free L\'evy-Schr\"odinger Hamiltonian}

To consider the properties of a free (without external potentials) L\'evy-Schr\"odinger semigroup, we  employ the
rescaled  Hamiltonians rather than semigroup generators that  have an opposite sign.
The pertinent  Hamiltonians have the form $\hat{H}=F(\hat{p})$, where  $\hat{p}=-i \nabla $ stands for the momentum  operator (up to the scaled away  $\hbar$ or $2mD$ factor), and  for $-\infty < k <+\infty $,  the function $F=F(k)$ is real valued and bounded from below. The action of $\exp(-t\hat{H})$ can be given  by means of an integral kernel $k_t\equiv
k(x-y,t)=k(y,0,x,t) $  where  $k_t(z,t)=\frac{1}{\sqrt {2\pi }}\int_{-\infty}^{\infty}\exp[-tF(p)+ipz]dp$.

Our further discussion is limited  to non - Gaussian random variables whose pdfs are centered and
symmetric, e.g. to  a subclass of stable distributions characterized by
\begin{equation}
F(p) = \lambda   |p|^{\mu } \, \,  \Rightarrow \hat{H}_{\mu }\, \equiv  \lambda
|\Delta |^{\mu /2} \, .
 \end{equation}
Here  $0<\mu <2$ and $\lambda >0$ stands for an intensity parameter
of  L\'{e}vy  process. To account for the interval  $0\leq \mu
\leq 2$ boundaries,  one should rather employ $(-\Delta )^{\mu /2}$
instead of $|\Delta |^{\mu /2}$;  $-\Delta $ is a positive operator.

The  pseudo-differential  Hamiltonian   $\hat{H}_{\mu }$,  by construction  is  positive and self-adjoint on a properly tailored  domain.  A  sufficient and necessary condition for both these  properties to hold true is that the pdf of the  L\'{e}vy process is symmetric, see Ref. \cite{applebaum}. The corresponding contractive semigroup admits an analytic continuation in time leading to L\'{e}vy-Schr\"{o}dinger equations and fractional quantum mechanics \cite{cufaro,laskin,laskin1}. The associated  jump-type dynamics is interpreted  in terms of L\'{e}vy flights. The pseudo-differential Fokker-Planck equation, which  corresponds to the  free fractional Hamiltonian $\hat{H}_{\mu }$ and the
fractional semigroup $\exp(-t\hat{H}_{\mu })=\exp(-\lambda |\Delta |^{\mu /2})$, reads
\begin{equation} \label{ffd}
\partial _t \rho  = -  \lambda |\Delta |^{\mu /2} \rho  \, ,
\end{equation}
to be compared with the ordinary heat equation $\partial _t \rho  = + D\Delta  \rho $.
 In particular $F(p)= \lambda  |p|$ refers to Cauchy process.

The action of the pseudo-differential operator $|\Delta |^{\mu /2}$ on a function can be expressed by the formula
\cite{klauder,olk}
\begin{equation} \label{tr}
 (|\Delta |^{\mu /2} f)(x)\, =\, - \int  [f(x+y) - f(x) ] \nu _{\mu }(dy),
\end{equation}
where $\nu _{\mu }(dy)$ is a corresponding L\'{e}vy measure (see, e.g. \cite{lev2}) and the integral in Eq. \eqref{tr} is understood in a sense of its Cauchy principal value. Changing  the integration variable $y$  to $z=x+y$ and employing
a definition of Riesz fractional derivative of the $\mu $-th order, \cite{dubkov}, we  arrive at
\begin{equation} \label{tr1}
 ( |\Delta |^{\mu /2} f)(x)\, =\, -  {\frac{\Gamma (\mu +1) \sin(\pi \mu/2)}{\pi }} \int  {\frac{f(z)- f(x)}{|z-x|^{1+\mu }}}\,
 dz \,
\end{equation}
with $( |\Delta |^{\mu /2} f)(x) =  -  \partial ^{\mu }f(x)/\partial |x|^{\mu }$.
The case of $\mu =1$  refers to the  Cauchy driver (e.g. noise).
We note a systematic sign difference between our notation for a pseudo-differential operator $|\Delta |^{\mu /2}$ and that
based on the fractional derivative notion, like e.g.  $ \Delta ^{\mu /2} \doteq     \partial ^{\mu }/\partial |x|^{\mu }$ of Refs.
\cite{sokolov,geisel}.

\subsection{Response to external potentials}

Consider now the L\'{e}vy-Schr\"{o}dinger Hamiltonian with external potential
\begin{equation}
\hat{H}_{\mu }  \equiv  \lambda |\Delta |^{\mu /2} +  {\cal{V}}(x)\, .
\end{equation}
Suitable  properties of ${\cal{V}}$ need to be assumed, so that $-\hat{H}_{\mu }$ is a legitimate  generator  of a
dynamical semigroup  $\exp(-t \hat{H}_{\mu })$, see e.g. Ref. \cite{olk}.

Looking for stationary solutions of the equation $\partial _t \Psi = \hat{H}_{\mu } \Psi$, we realize that
if a square root of a  positive  invariant   pdf  $\Psi \sim \rho _*^{1/2}$ is asymptotically  to  come out, then
fractional Sturm-Liouville operator should be used to derive an explicit form of  $\rho _*^{1/2}$ for a given ${\cal{V}}$.
In the opposite situation, when $\rho _*(x)$ is a priori prescribed, we can determine ${\cal{V}}$ through a compatibility condition:
\begin{equation} \label{calv}
{\cal{V}}  =   -\lambda\,  {\frac{|\Delta |^{\mu /2}  \rho ^{1/2}_*}{\rho ^{1/2}_*}} \, .
\end{equation}
The main point here is that we do not have too much freedom in pre-selecting  a functional form of  $\rho _*$, as
suitable conditions need to be respected by the inferred auxiliary potential ${\cal{V}}$,   to yield a
contractive semigroup dynamics. This leads to a conclusion \cite{olk}  that only under the contractive  semigroup premises,  an invariant pdf
of a jump-type process may actually become its asymptotic target.  The detailed  discussion of this issue for Cauchy semigroups can be
found in Ref. \cite{olk}.

To derive  the pseudo-differential equation governing the behavior of a system in an external potential, we rewrite the pdf of the semigroup-driven stochastic process in the form \eqref{un1} (see also \cite{stef}).  Any strictly  positive function (here we consider only functions that vanish for large $x$) can be rewritten in an exponential form. Hence,
by adopting the notation $\rho _*(x) = \exp [2\Phi (x)]$ and  accounting for \eqref{calv} we arrive at a continuity equation with an
explicit fractional input
 \begin{equation} \label{ptr}
  \partial _t \rho  =   -   \lambda  ( \exp \Phi ) |\Delta |^{\mu /2}[ \exp(-\Phi )
    \rho ] +   {\cal{V}} \,  \rho  \, .
 \end{equation}
 The definition \eqref{calv} suggests that \it  any \rm  pdf of the form $\rho _*(x) =\exp [2\Phi (x)]$
 is   a proper candidate for  a stationary solution of the Eq. \eqref{ptr}. However things are not that simple.
 It is  only  the  semigroup dynamics generated by
\eqref{calv}  and  \eqref{ptr}  that may guarantee a consistent temporal approach towards  an asymptotic invariant density of the
 stochastic process in question.  Right at this point, an issue  of  restrictions upon an effective potential ${\cal{V}}$ of  Eq.
  \eqref{ptr},  \cite{olk,stef}, enters the stage.

It is instructive to mention that for  L\'{e}vy flights in external
force fields, the  (somewhat left aside) Langevin approach is
 known to yield \cite{fogedby} a continuity (e.g. fractional Fokker-Planck) equation in a very different form
\begin{equation} \label{dby}
\partial _t\rho = -\nabla \left(- {\frac{\nabla V }{m\beta }}\, \rho  \right) - \lambda |\Delta |^{\mu /2}\rho.
 \end{equation}
Even  if we know \cite{stef}, that   an asymptotic  invariant
density of Eq. \eqref{ptr} may coincide with that for Eq.
\eqref{dby}, these two transport equations refer to different
temporal patterns of behavior.

We note that, in contrast to the semigroup modeling,   the Langevin scenario for L\'{e}vy flights in confining potentials has received  ample attention in the literature, see  \cite{klafter,fogedby,dubkov,chechkin,chechkin1} to cite a few.

\subsection{Topologically induced super-diffusions}

 It is of some interest to  invoke an independent (so-called topological, see above) approach of Refs. \cite{sokolov,geisel} where one modifies jumping rates by suitable local factors  to arrive  at a response  mechanism that is characteristic
of the previously outlined semigroup dynamics. Namely, in view of (9), the free transport equation (7) can be  re-written as a master
equation:
\begin{equation}
\partial _t \rho (x)  = \int [w(x|z) \rho (z) - w(z|x) \rho (x)] \nu _{\mu }(dz)\, .
\end{equation}
 The jump rate is  an even function,
$w(x|z)= w(x|z)$.
However, if we replace  the  jump rate
\begin{equation} \label{unn1}
w(x|y)\sim 1/|x-y|^{1+\mu }
\end{equation}
(c.f. Eq. \eqref{tr1}) by  the expression
\begin{equation} \label{unn2}
w_{\phi }(x|y)\sim {\frac{\exp [\Phi (x) - \Phi (y)]}{|x-y|^{1+\mu }}}
\end{equation}
and
account for the fact that $w_{\phi }(x|z) \neq w_{\phi }(z|x)$,   the corresponding transport equation  takes the form:
\begin{eqnarray}
&&(1/\lambda )\partial _t \rho = |\Delta |^{\mu /2}_{\Phi }f =
-  \exp (\Phi )\, |\Delta |^{\mu /2}[ \exp(-\Phi ) \rho ]   +\nonumber \\
&& +\rho \exp (-\Phi ) |\Delta |^{\mu /2} \exp(\Phi ) \, .\label{unn3}
\end{eqnarray}
Whatever potential $\Phi (x) $ has been chosen (up to a normalization factor), then  formally
 $\rho _*(x) =\exp (2\Phi (x))$ is a stationary solution of   Eq. \eqref{unn3}.
Moreover, one readily verifies that Eq.  \eqref{unn3}  is identical with the semigroup-induced   Eq.  \eqref{ptr}.

Accordingly, if for a pre-determined  $\rho _*=\exp(2\Phi )$, there exists the semigroup potential   ${\cal{V}}$,
Eq.  \eqref{calv}, then   Eq. \eqref{unn3} defines the dynamics that  belongs to  the previously
outlined semigroup framework. This entails a direct verification of whether  the stationary density $\rho _*$
may really  be interpreted as an asymptotic target of the pertinent fractional transport equation.

Stationary pdfs for the topologically-induced dynamics  were demanded  to occur  in  a Gibbsian form $\exp (2\Phi)$, see  \cite{sokolov,geisel} for a possible phenomenological background for this assumption. Therefore,  we redefine  $\Phi$ as follows. Rewriting the stationary pdf $\rho _*$ as  $\rho _*(x) =(1/Z) \exp(-V_*(x)/k_BT)$ (normalization factor $Z$ stands for a partition function), we recover  a function $V_*(x)= -  k_BT \ln (Z\, \rho _*(x))$ that receives an interpretation of the  external potential.
With these re-definitions, the previous equation \eqref{unn3} takes the customary form employed in the discussion of topologically induced super-diffusions:
\begin{widetext}
\begin{equation}
\partial _t \rho =-    \exp(-\kappa V_*/2)\,  |\Delta |^{\mu /2}
\exp(\kappa V_*/2 )  \rho  + \rho \exp (\kappa V_*/2) |\Delta |^{\mu /2}
\exp(-\kappa V_*/2),\ \kappa = 1/k_BT. \label{fk}
\end{equation}
\end{widetext}

\section{Response of Cauchy noise to  polynomial  potentials}

\subsection{Ornstein - Uhlenbeck - Cauchy process}

In case of the Ornstein - Uhlenbeck - Cauchy (OUC) process, the
drift is given by $b(x)= - \gamma x$, and an asymptotic invariant
pdf associated with
\begin{equation}\label{zel}
\partial _t \rho = - \lambda |\nabla | \rho + \nabla [(\gamma x)\rho ]
\end{equation}
 reads:
\begin{equation} \label{couc}
\rho _*(x) = {\frac{\sigma }{\pi (\sigma ^2 + x^2)}},
\end{equation}
where $\sigma = \lambda /\gamma $, c.f. Eq. (9) in Ref. \cite{olk1}.

Note that a characteristic function of this density reads $F(p) = -
\sigma  |p|$ and gives account for a  non - thermal
fluctuation - dissipation balance. The modified noise intensity
parameter $\sigma $ is  a ratio of an intensity parameter  $\lambda$
of the  Cauchy noise and  of the friction coefficient $\gamma $.

The invariant density of OUC process \eqref{couc} generates (with
the help of Eq. \eqref{calv}) the  following Cauchy semigroup
potential ${\cal{V}}$ (we set $\mu =1$)
\begin{eqnarray}
&&{\cal{V}}(x) = {\frac{\lambda }{\pi }} \left[ - {\frac{2}{\sqrt{a}}}
+ {\frac{x}{a}}\ln {\frac{\sqrt{a}+x}{\sqrt{a}-x}}\right], \label{oucpot} \\
&&a=\sigma ^2 + x^2.\nonumber
\end{eqnarray}
The potential \eqref{oucpot} had been analyzed  in Ref. \cite{stef}.
It is clear that ${\cal{V}}(x)$  is bounded both from below and
above, well fitting to the general mathematical construction  of
(semigroup-driven or topological) Cauchy processes in external
potentials, \cite{olk}.

\begin{figure}
\begin{center}
\includegraphics [width=0.9\columnwidth]{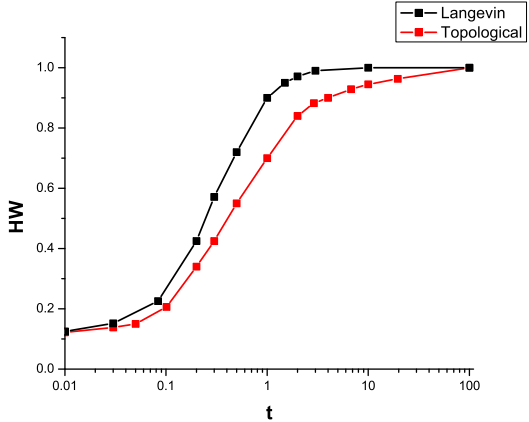}
\end{center}
\caption{Temporal behavior of the half-maximum width (HW): for the OUC process in Langevin-driven and semigroup-driven (topological) processes. Motions begin from common initial data $\rho(x,t=0)=\delta(x)$ and end up at a common pdf \eqref{couc} for $\sigma=1$.}\label{fig:hw}
\end{figure}

Since the OUC pdf \eqref{couc} has no variance, in Fig. \ref{fig:hw} we visualize the temporal evolution of OUC process with initial data localized at $x=0$ ($\rho(x,t=0)=\delta(x)$) in two motion scenarios (i.e. Langevin and semigroup driven) by comparing the width of the OUC "bell" at its half-maximum at a number of consecutive instants of time. It is seen that Langevin dynamics sets at equilibrium faster than the semigroup-induced dynamics.

\begin{figure}
\begin{center}
\includegraphics [width=0.9\columnwidth]{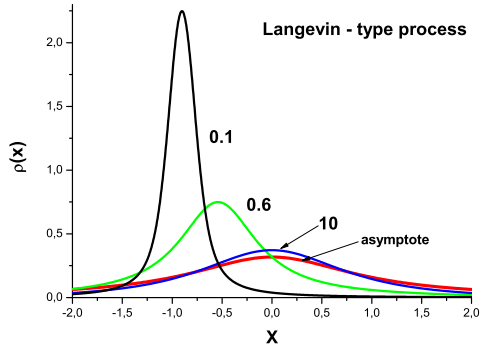}
\end{center}
\caption{Time evolution of Langevin-driven pdf $\rho_L(x,t)$ beginning from the initial data $\rho_L(x,t=0)=\delta(x+1)$ and ending at the pdf \eqref{couc} (shown as "asymptote" in the figure) for $\sigma=1$. Figures near curves correspond to $t$ values. }\label{fig:hw1}
\end{figure}

Further check of the OUC dynamics is to obtain (numerically \cite{ux}) the time evolution $\rho(x,t)$ of  initial data, localized in another point.
Since quadratic potential (corresponding to OUC process) has a minimum at $x=0$, we investigate the temporal evolution for the initial data   that are shifted from the point $x=0$. Namely, we consider
two cases - "left" $\rho_L(x,t=0)=\delta(x+1)$ and "right" $\rho_R(x,t=0)=\delta(x-1)$ ones. As the "left" and "right" cases are symmetric with respect to $y$ axis, in Fig.  \ref{fig:hw1} we report the "left" case only.
It is seen that final stage of the evolution is still the invariant pdf \eqref{couc}. This means that fictitious particle representing our
process, "rolls down" to potential minimum at $x=0$. We note that this "rolling down" occurs slower then the evolution with initial data,
 localized at $x=0$. The topologically-induced dynamics of $\rho_{L,R}(x,t=0)$ is qualitatively the
 same, except  for the just mentioned slow-down,
 as Langevin-induced one (see Ref. \cite{stef}) and we do not show it here.

A pre-selection of the OUC $\rho _*$, Eq. \eqref{couc},  proved to provide (through a numerical process reconstruction)  a consistent asymptotic target pdf for the Cauchy semigroup-driven dynamics. At this point we may justifiably ask about  limitations  upon  freedom of choice that is  admitted in such  invariant pdf pre-selection procedure.

{\it Remark 1:} As a byproduct of the discussion we have established
a pseudo-differential Hamiltonian system, whose  ground state equals
the square root  $\rho _*^{1/2}(x)$ of the Cauchy pdf \eqref{couc}.

\subsection{Cauchy driver: Polynomial drifts  and the semigroup evolution}

The OUC case  corresponds to a linear drift function.  A number of polynomial drift functions has been discussed
in the literature with a focus on confining features of various external forces on L\'{e}vy flights. In each case a corresponding asymptotic invariant pdf has been found, albeit with a restriction  (in view of limited computational facilities) to the Cauchy driver. For  clarity of subsequent discussion below we provide  some  examples.

The quadratic Cauchy density $ \rho _*(x) = 2/{\pi}(1+x^2)^2$ stands for an asymptotic target of the Langevin - driven
process with the drift  $b(x) =(-\gamma x/8)(x^2+3)$ \cite{stef}. For the above quadratic Cauchy pdf, the  associated semigroup evolution is defined, by means of the potential function ${\cal{V}} (x)  = \lambda
(x^2-1)/(x^2 +1)$, obtained from \eqref{calv}. This potential is also bounded from below and above and consequently, \cite{olk}, defines a contractive semigroup generator and hence semigroup dynamics.

\begin{figure}
\begin{center}
\includegraphics [width=0.9\columnwidth]{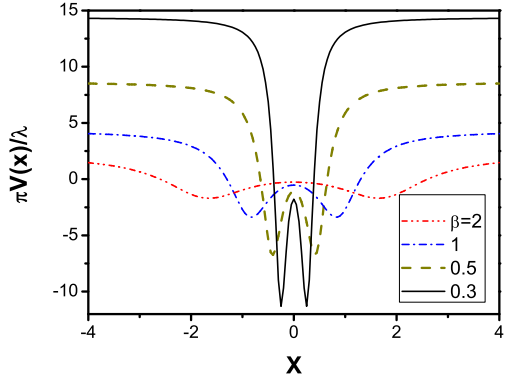}
\end{center}
\caption{The coordinate dependence of ${\cal{V}}(x)$ associated with the bimodal pdf,
 for different values of   $\beta$. } \label{fig:zz}
\end{figure}

 The bimodal pdf $\rho_*(x)=\beta^3  / \pi (x^4-\beta^2x^2+\beta^4)$ can be derived from the drift function $b(x)=-\gamma x^3/ \beta^3$.  The associated semigroup dynamics is correctly defined by means of a potential function ${\cal{V}}$, obtained from Eq. \eqref{calv} numerically.  Its shape is reported in Fig. \ref{fig:zz}. It is seen, that the potential is
 bounded from below and above, again perfectly fitting to the general theory of \cite{olk}.

\subsection{Cauchy driver: Gibbsian versus non-Gibbsian asymptotics}

The Langevin-driven dynamics with a given drift  $b\sim -\nabla V$, where $-\nabla V$
stands for  a conservative force  acting upon a particle in the course of its random motion,
for non-Gaussian driver (like e.g. Cauchy one) does not admit an asymptotic invariant density in the Gibbsian form $\rho _* \sim \exp (-V/k_BT)$, \cite{klafter}.

As we have established above, see also \cite{stef},  the Langevin-driven and semigroup-driven jump-type processes, with the Cauchy driver, may share common asymptotic stationary pdfs, that are obviously non-Gibbsian.

On the other hand, as mentioned in Section II.D, one may suspect
that  asymptotic invariant pdfs of a semigroup-driven process (in
other words, topologically-induced one) may not necessarily have fat
(e.g. non-Gaussian) tails, due to extremely strong confinement   of
admissible jumps,  imposed by the "potential  landscape" of an
inhomogeneous medium. That amounts to assuming that a Gibbs-type
asymptotic pdf $\rho _* =(1/Z) \exp(-V_*/k_BT)$  may be employed in
the construction of the topologically-induced dynamics \eqref{fk}.

A  thermalization  mechanism, under which the equilibrium conditions could have been achieved for such Gibbsian
$\rho _*$ (the non-Gaussian mechanism is excluded \cite{klafter} and has  not been considered in
  Refs. \cite{sokolov,geisel}) is actually unclear. The major focus in the literature was upon a suitable "potential landscape" (potential profile), such that local modifications \eqref{unn2} of the jump rates would drive the random motion towards a Gibbsian equilibrium.

We shall follow the "potential landscape" intuition and  refer to explicit  dynamics simulations reported in Ref. \cite{sokolov}
for  the double well potential $V_*(x)\equiv \Phi(x)= x^4 -2x^2 +1$.  Dimensional units are scaled away and, to facilitate comparison,
the notation $\Phi $ refers presently to that of Ref. \cite{sokolov}), see Fig. 3 therein.
As a  supplementary  test, we consider also $\Phi \equiv V_*(x)= x^2$,  leading to Gibbsian asymptotic pdf in the Gaussian form.

\begin{figure}
\begin{center}
\includegraphics [width=0.9\columnwidth]{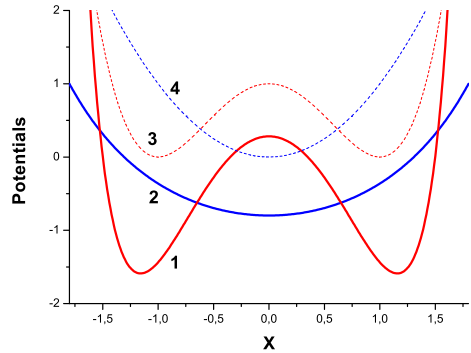}
\end{center}
\caption{The coordinate dependence of the semigroup potential ${\cal{V}}(x)$ (curves 1 and 2), corresponding to $V_*(x)= x^4 -2x^2 +1$ (curve 3) and $ V_*(x)= x^2$ (curve 4), respectively. Curves  3 and 4 are shown for a comparison with, strikingly similar in shape, semigroup potential curves  1 and 2}\label{fig:zz1}
\end{figure}

Our next step was to evaluate (numerically) the semigroup potential
${\cal{V}}$, after literally  substituting a concrete  Gibbsian
asymptotic pdf $\rho _*$ to Eq. \eqref{calv}. The outcome is
reported in Fig. \ref{fig:zz1}.

The semigroup potentials ${\cal{V}}$  depicted in Fig. \ref{fig:zz1}  are  admitted by the general theory of Ref.  \cite{olk}
 (see below for details), so granting the operator $\exp(-t\hat{H})$ ($\hat{H}= - |\nabla | - {\cal{V}}$),
   the contractive semigroup status. This  proves that  an invariant (stationary) density of Eq. \eqref{zel},
  with the bimodal or harmonic exponent, actually is an {\it asymptotic invariant} pdf of the topological process.

  The topological confining  mechanism appears to be much stronger than confining mechanism based on the Langevin modelling.
  Namely, if both mechanisms share the same potential $\Phi(x)=V_*(x)$, then the asymptotic pdf in Langevin scenario has
  no more then a finite number of moments. At the same time, the corresponding asymptotic pdf of the topological process,
   being in the form $\propto \exp(-\Phi)$, under the Gibbsian premises may in principle  admit all moments.
   To the contrary, if we impose for both mechanisms to
   have the same asymptotic pdf, the above statement is invalid.

For completeness, we now recollect the above mentioned  requirements upon ${\cal{V}}$ that need to be observed in the presence of Cauchy driver, \cite{olk}. Namely, the potential should allow to be made positive (by merely adding a  constant), should be locally bounded and needs to be measurable (i.e. can be approximated with arbitrary precision by step functions sequences). The  Cauchy generator plus a potential  with such properties is known to  determine uniquely \cite{olk} an associated Markov process of the jump-type and its step functions approximation.  The limiting behavior of the pertinent step process, as  the step size is going to zero, remains under control.

{\it Remark 2:} Would we have followed the standard Langevin
modeling for the Cauchy driver, with the  external force potential
$V_*(x)= x^4 -2x^2 +1$ and the resultant drift $-\nabla V_* = b$, an
invariant pdf of the corresponding fractional Fokker-Planck equation
would have the form:
  \begin{equation}\label{zuz}
  \rho _*(x) = {\frac{2a(a^2 +b^2)}{\pi }} {\frac{1}{(a^2+b^2)^2 + 2(a^2 -b^2)x^2 + x^4}}
  \end{equation}
with $a\simeq 0.118366$ and $b\simeq 1.0208$. Here,
$a$ and $b$ are, respectively, real and imaginary parts of complex roots of the  cubic equation  $z^3 +z - 1/4=0$.
It is easy  to show both analytically and numerically that  the above $\rho _*$ is properly normalized.
 By using the formula \eqref{calv} we may associate with  the non-Gibbsian pdf a semigroup potential.
  Its properties prove that we deal with a contractive  semigroup dynamics and a common asymptotic pdf for both Langevin and
  semigroup motion scenarios.  The outcome of this discussion is depicted in Fig 5.
\begin{figure}
\begin{center}
\includegraphics [width=0.9\columnwidth]{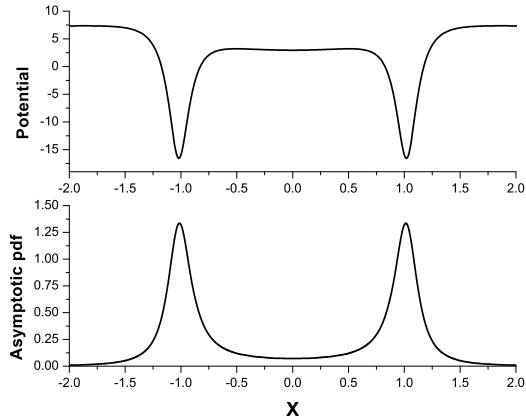}
\end{center}
\caption{The coordinate dependence of the semigroup potential
${\cal{V}}(x)$ derived from the non-Gibbsian pdf  of  Remark 2,
Section III.C, along with the pdf \eqref{zuz}.}
\end{figure}
\\
{\it Remark 3:} Standard arguments , c.f.  Section II.A, convince us
that the Gibbsian form  $\rho _* \sim \exp(-V_*/k_BT)$  of an
asymptotic density would have been recovered  in the presence of the
Wiener (Brownian) noise, given the drift function  $b\sim - \nabla
V_*$ with $V_*(x)= x^4 -2x^2 +1$ or $V_*(x)= x^2$.

\subsection{Cauchy oscillator}
Our preceding discussion has  actually been  related  to the
generalization of the reverse engineering problem of Ref.
\cite{klafter}:  (i) with an invariant  pdf $\rho _*(x)$  in hands,
derive $b(x)$ for the associated Langevin equation to reconstruct
the Langevin-drien dynamics of the pdf, (ii) from the  same
invariant pdf $\rho _*(x)$, deduce the topologically - driven
potential ${\cal{V}}(x)$ for the related Cauchy semigroup dynamics
(thus granting that pdf an asymptotic pdf status).

In the present section we invert the reasoning and effectively
follow the direct engineering route: having ${\cal{V}}(x)$, we
 employ the semigroup dynamics principles to infer an asymptotic
target $\rho _*(x)$. This construction will be supplemented by
identifying, if any,  the drift function $b(x)$ for the associated
Langevin process that gives rise to  the same asymptotic target.

To  this end it seems natural to resort to simplest possible
potential functions. Our choice (motivated by a simplicity of
involved Fourier transforms and an immediate comparison with the
Gaussian OU process of Section II.A) is the  familiar harmonic
oscillator potential $ {\cal{V}}(x)= {\frac{\kappa }2} \,  x^2  -
{\cal{V}}_0, $ constant ${\cal{V}}_0$ is  left unspecified.

Our major object of interest is thus a pseudo-differential Hamiltonian (this is a very special, massless version, of the well known Hamiltonian
for the relativistic harmonic oscillator problem, \cite{klauder,applebaum}):
\begin{equation} \label{rel1}
\hat{H}_{1/2 }  \equiv  \lambda |\nabla |  + \left( {\frac{\kappa }2} \,  x^2  - {\cal{V}}_0\right) \, .
\end{equation}
That in turn  is a Cauchy analog of the familiar harmonic oscillator Hamiltonian of Section II.A:
\begin{equation} \label{rel2}
\hat{H} = -D\Delta + \left( {\frac{\gamma ^2x^2}{4D}}  - {\frac{\gamma }2} \right) \, .
\end{equation}
Since the quadratic function is admissible \cite{olk} as a   semigroup  potential,
 we need not to bother about an asymptotic approach towards an invariant density, but follow a direct reconstruction route: given ${\cal{V}}$ of  Eq. \eqref{rel2},  deduce an invariant  pdf $\rho _*$.
To this end we turn back to Eq. \eqref{calv} with $\mu =1$ and consider a pseudo-differential equation
 \begin{equation}\label{rel3}
\left({\frac{\kappa }2} \,  x^2  - {\cal{V}}_0\right) \rho ^{1/2}_* =   -\lambda\,  |\nabla |\,  \rho ^{1/2}_*
\end{equation}
to solve it   with respect to $\rho _*$.

We denote  $\tilde{f}(p)$ the Fourier transform of $f=\rho _*^{1/2}(x)$.
Accordingly, Eq. \eqref{rel3} takes the following form:
\begin{equation}\label{rai1}
-{\frac{\kappa }2} \Delta _p \tilde{f} +  \gamma |p| \tilde{f}= {\cal{V}}_0 \tilde{f}
\end{equation}
which, up to constants adjustments,  can be recognized as  the eigenvalue problem for the  Schr\"{o}dinger operator with   the linear confining  (modulus of the argument) potential, \cite{robinett}, see also \cite{ruiz,moreno,landau}.
 Albeit  with  respect to momentum - space  (here, wave-vector) variables, i.e. with $\Delta _p= d ^2/d p^2$
replacing the conventional spatial Laplacian.

\begin{figure}
\begin{center}
\includegraphics [width=0.9\columnwidth]{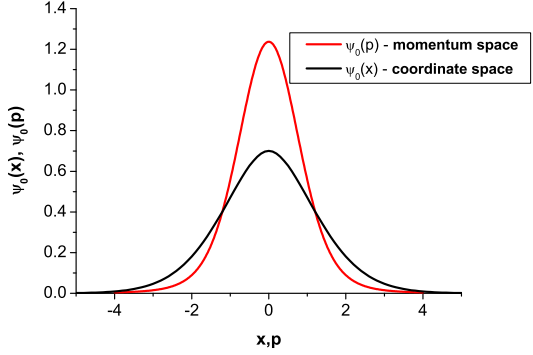}
\end{center}
\caption{Fourier image $\psi _0(k)$   of the  function  $\psi _0(x)$ along with this
function}\label{Fig:fur}
\end{figure}

By changing an independent variable $p$ to $k= (p-\sigma)/\zeta$, next denoting $\psi (k) = \tilde{f}(p)$ with the  identifications $\sigma = {\cal{V}}_0/\gamma $ and $\zeta = (\kappa /2\gamma )^{1/3}$, we may rewrite the above eigenvalue problem (with ${\cal{V}}_0$ standing for an eigenvalue) in the form of the following ordinary differential  equation
\begin{equation} \label{rai2}
{\frac{d^2 \psi (k)}{dk^2}} = |k| \psi (k),
\end{equation}
whose solutions can be represented in terms of Airy functions. A brief resume of how to deduce the eigenfunctions and eigenvalues of the  original   problem \eqref{rai1}, is  relegated to Appendix.
A  unique normalized ground state function  of the problem \eqref{rai1}, \eqref{rai2} is composed of two Airy pieces that
are glued together at the first zero $y_0$ of the Airy function derivative:
\begin{eqnarray}\label{frd17}
&&    \psi _0(k)=A_0\left\{
\begin{array}{c}
{\rm Ai}(-y_0+k),\ k>0 \\
{\rm {Ai}}(-y_0-k),\ k<0,
\end{array}
\right.\nonumber \\
 &&A_0=\left[ \mathrm{Ai}(-y_0)\sqrt{2y_0}\right] ^{-1}, \ y_0 \approx
 1.01879297.
\end{eqnarray}
The function \eqref{frd17} is  a square root of the pdf in momentum
space. We reproduce its shape in Fig. \ref{Fig:fur}.

To transform the  ground state  solution  back to coordinate space, we evaluate the inverse Fourier transformation of the ground state solution \eqref{frd17}, see Appendix C for details. This yields the following real ground state wave function $f(x) \to \psi _0(x)$
\begin{equation}\label{aif6}
\psi _0(x)=\frac{A_0}{\pi}\int_{-y_0}^\infty{\rm Ai}(t)\cos x(t+y_0)dt =
\rho _*^{1/2}(x),
\end{equation}
which determines an invariant pdf $\rho _*(x)$ of the direct engineering problem \eqref{rel3} as follows:
\begin{eqnarray}\label{aif7a}
&&\rho_*(x)=\left(\frac{A_0}{\pi}\right)^2\left[\int_{-y_0}^\infty{\rm Ai}(t)\cos
    x(t+y_0)dt\right]^2  \\
&&\equiv \left(\frac{A_0}{\pi}\right)^2\int_{-y_0}^\infty dt \int_{-y_0}^\infty dt_1
 {\rm Ai}(t){\rm Ai}(t_1) \times \nonumber \\
 &&\times \cos x(t+y_0)\cos x(t_1+y_0).  \nonumber
\end{eqnarray}
\begin{figure}
\begin{center}
\includegraphics [width=0.9\columnwidth]{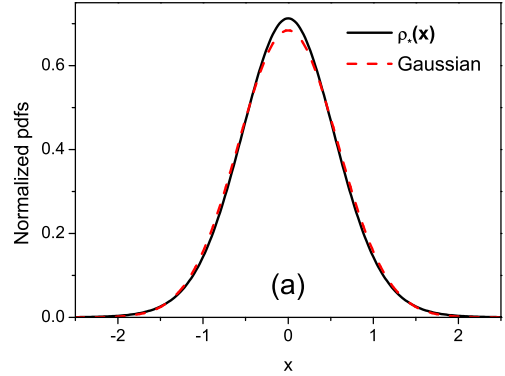}
\includegraphics [width=0.9\columnwidth]{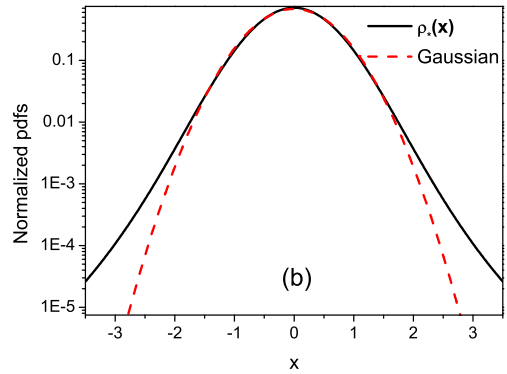}
\end{center}
\caption{Normalized invariant pdf \eqref{aif7a} (full
line) for the quadratic semigroup potential.
The Gaussian function, centered at $x=0$ and with  the same variance
$\sigma ^2=0.339598$ is shown for comparison. Panel (a) shows
functions in linear scale, while panel (b) shows them in logarithmic
scale to better visualize their different behavior.}\label{fig:f1}
\end{figure}

Even with an exact analytic formula for the normalized $\int_{-\infty}^\infty \rho_*(x)dx=1$ function $\rho_*(x)$
\eqref{aif7a}  in hands, a better insight into its properties is achieved only   by  means of numerical methods.
 We depict both $x$ and $p$-space versions of $\psi _0$ in Fig.6.

In Fig.7, the resultant invariant density $\rho _*(x)$ is
represented by a full line. For comparison,  we have depicted  the
Gaussian function with the same variance $\sigma ^2=0.339598$,
$f(x)=\frac{1}{\sigma\sqrt{2\pi}}e^{-\frac{x^2}{2\sigma ^2}}$. It is
shown by a dashed line.  Fig. \ref{fig:f1}b shows both above
functions in the  log scale. It is clearly seen, that Airy-induced
$\rho_*(x)$ decays at infinities slower then Gaussian.

Let us finally note, that the potential $\Phi (x)$ may be recovered from the ground state function \eqref{aif7a} as the asymptotic pdf  for topologically driven (semigroup) process has the form $\exp(-\Phi)$. Namely, $\Phi(x) \propto -\ln \rho_*(x)$, where $\rho_*(x)$ is given by Eq. \eqref{aif7a}. The inversion of the $y$ axis of Fig. \ref{fig:f1}b shows this potential (compared with harmonic one $y=x^2$).

\subsection{Reverse engineering for the Cauchy oscillator ground state pdf}

\begin{figure}
\begin{center}
\includegraphics [width=0.9\columnwidth]{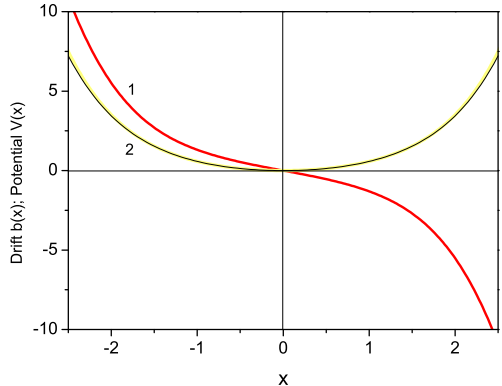}
\end{center}
\caption{Langevin - type drift $b(x)$ (curve 1) and its (force) potential $V(x)$
(curve 2), that  give rise  to an invariant density (30).\label{fig:f2}}
\end{figure}

For a given $\rho _*$ the definition of a drift function $b(x)$ (we put either $\lambda
=1$ or define $b \to b/\lambda$) is:
\begin{equation}\label{dr1}
    b(x)=-\frac{1}{\rho_*(x)}\int
    [|\nabla|\rho_*(x)]dx\equiv
\end{equation}
$$
    \frac{1}{\pi \rho_*(x)}\int dx \int _{-\infty}^\infty
    \frac{\rho_*(x+y)-\rho_*(x)}{y^2}dy\, .
$$
Inserting $\rho _*(x)$, Eq. \eqref{aif7a}, we  get
\begin{equation}\label{dr9}
    b(x)= -\frac{\int_{-y_0}^\infty {\rm Ai}(t)\sin
    x(t+y_0)dt}{\int_{-y_0}^\infty {\rm Ai}(t)\cos
    x(t+y_0)dt}.
\end{equation}
The final formula for $b(x)$, \eqref{dr9}, together with that for the  corresponding
Langevin  force potential $V(x)\equiv -\int b(x)dx$  look  a bit clumsy. Therefore it is
appropriate to reiterate to numerics, see  Fig. \ref{fig:f2}.
The plot of $b(x)$  is reported in Fig.\ref{fig:f2} along with
potential function $V(x)\equiv -\int b(x)dx$.

We were unable to determine the asymptotic of $b(x)$ at spatial  infinities.
It is possible, however, to expand Eq. \eqref{dr9} in power series at small $x$. It turns out, that these
truncated series describe the function $b(x)$ (and correspondingly $V(x)$) surprisingly well.

To obtain this approximation, we expand both numerator and
denominator of Eq. \eqref{dr9} in power series to obtain $
\int_{-y_0}^\infty {\rm Ai}(t)\sin
    x(t+y_0)dt=0.824278x-0.503237 x^3+0.205648 x^5-0.0650381 x^7+...
$  for numerator and
$ \int_{-y_0}^\infty {\rm Ai}(t)\cos
    x(t+y_0)dt=0.809073-0.687715 x^2+0.334243 x^4-0.118792
    x^6+0.0339885 x^8 -...$
for denominator. The series can be easily continued for larger amount of terms. Now, the ratio of these series can be also
expressed in the form of the  series. Here, we reproduce  only the truncated (up to $x^5$) version of the series
\begin{eqnarray}\label{cns3}
    b(x)&\approx& -1.01879x-0.243986x^3-0.04056 x^5,\nonumber \\
    V(x)&\approx&
    0.509395 x^2+0.0609965 x^4+0.00676 x^6.
\end{eqnarray}
The approximations \eqref{cns3} can be used to calculate numerically the Langevin-type dynamics with
the  invariant density \eqref{aif7a}.

\section{Conclusions}

 The L\'{e}vy - Schr\"odinger semigroup modeling sets a  (mathematically rigourous) link between L\'{e}vy flights and pseudo - differential
  Hamiltonian systems. It is well known that the pdf of a free L\'{e}vy flight has so-called heavy tails and  consequently does not have second
  and higher moments. Properly tailored  external inhomogeneities are capable of  "taming" a L\'{e}vy flight so that the resultant
  pdf admits higher moments which, in turn, makes possible to use such functions for the description of real physical (and
other,  like biological or economic) systems.

We have encoded the overall  impact of inhomogeneities in the
semigroup potential notion, that may be interpreted as an external
potential in the affiliated pseudo-differential Hamiltonian
operator.  However, not any conceivable  semigroup potential
 and thus, not any conceivable  inhomogeneity, can  make the corresponding  jump-type  process  a
  mathematically well-behaved  construction,   where   the  Markovian dynamics entails  an  approach  towards   a  unique  stationary pdf.

In the present paper, we have investigated  the behavior of L\'{e}vy
oscillators in different external confining potentials. Our analysis
shows that to control the long - time asymptotics of the above
L\'{e}vy oscillator pdfs,  suitable restrictions upon the
Hamiltonian and its (semigroup or topologically-induced)  potential
${\cal V}(x)$ need to be observed, \cite{olk}.

Namely,  ${\cal V}(x)$ should allow to be made positive (this is
achieved by simple vertical shift of the entire function), should be
locally bounded and needs to be measurable, i.e. should have a
possibility to be approximated with arbitrary precision by step
functions sequences. The fulfilment of these requirements provides
verifiable criteria for an invariant pdf to be actually a time -
asymptotic pdf of a semigroup (equivalently, topologically)-driven
process.

A technical   advantage of a semigroup formalism is the possibility
of eigenfunction expansions  for $\hat{H}$ which allows to deduce
explicit formulas for transition pdfs,  \cite{risken}.  As a
byproduct of the above procedure we have completely solved the
eigenvalue problem for the Hamiltonian operator of so-called Cauchy
oscillator in terms of Airy functions. The ground state wave
function of such oscillator has been obtained analytically. Its
square defines the invariant pdf of Cauchy oscillator process. The
latter pdf is approached by the jump-type process at large times.

We have extended the targeted stochasticity problem of Ref.
~\cite{klafter} to  the  above semigroup-driven  (topological)
L\'{e}vy processes, which are widely used in literature to model
various  systems, like polymers, glasses and complex networks. Our
departure point was  as follows: having an invariant pdf $\rho
_*(x)$,  recover not only the   Langevin  drift $b(x)$ and potential
$V(x)=-\int b(x)dx$, but also the potential ${\cal V}(x)$ of the
corresponding topological (semigroup) L\'{e}vy process, being
attributed to the same invariant pdf.

Furthermore,  we have relaxed a common pdf requirement and have
reformulated the targeted stochasticity problem as a task of
reproducing a suitable contractive semigroup, given an invariant
pdf, with  the L\'{e}vy (specifically, Cauchy) driver in action. We
have shown, that the semigroup modeling provides much stronger
confining properties than the standard Langevin one, such that the
resultant asymptotic pdf may have all moments.

To be more specific, if both above approaches  involve (albeit
differently)  the same conservative force  potential $\Phi =
V_*(x)$, then the asymptotic pdf in the  Langevin scenario, being an
inverse polynomial, has no more then a finite number (the degree of
the polynomial minus 2) of moments. At the same time, the
corresponding asymptotic pdf of the topological process, being of
the Gibbs  form $\propto \exp(-\Phi)$, may in principle admit all
moments. If both mechanisms  refer to a common  asymptotic pdf, the
latter being derivable  in the  Langevin approach, the previous
statement is no longer valid, \cite{stef}.

It turns out that the  asymptotic behavior of a time-dependent pdf,
in the semigroup (topological) modeling, may critically depend on
the initial data choice, like e.g. the location  of the initial  pdf
in the "potential landscape" of $V_*(x)$.  The signature of such
behavior is encoded in the functional form of a
 semigroup potential ${\cal V}(x)$, derived from  the a priori chosen
  invariant pdf $\rho _*(x)$,
   by means of above generalized reverse engineering procedure.
  If the effective (semigroup) potential obeys the requirements of
   \cite{olk},   we may expect that  a prescribed invariant density  is indeed approached
   in  the large time asymptotic of the random process, irrespective of
   the initial  (pdf) data choice.  These  requirements need to be
   verified for each specific guess about a functional form of  the prescribed  invariant pdf $\rho _*$.

\appendix

\section{Schr\"{o}dinger eigenvalue problem for the linear potential and Airy function}

In the present Appendix we briefly recapitulate, basically
retrievable  in the literature but not accessible in minute detail
nor in a closed form,  (see, e.g. \cite{robinett}), a procedure of
construction of the eigenfunctions of equation \eqref{rai2}.
 Symmetry arguments (see, e.g. \cite{robinett}) and  an  explicit solution of corresponding Schr\"{o}dinger
 equation \cite{landau} lead to the following expression for the eigenfunctions (here we substitute $y$ for $k$)
\begin{equation} \label{spe1}
\psi _n(y)=\left\{
\begin{array}{c}
A_n\ {\rm Ai}(-y_n+y),\ y>0 \\
\pm A_n\ {\rm {Ai}}(-y_n-y),\ y<0,
\end{array}
\right.
\end{equation}
where $n$ enumerates eigenvalues (and corresponding
eigenfunctions). In Eq. \eqref{spe1}, $y_n$ numbers the n-th zero of the function Ai$(y)$ (for odd
$n$), or  of  its derivative for even $n$ (observe that zeros of both
function Ai and its derivative lie on the negative semi-axis).  $A_n$
is a normalization coefficient, determined by the standard identity
\begin{equation}\label{spe2}
    A_n^2\int_{-\infty}^\infty \psi _n^2(y)dy=1.
\end{equation}
Here we use simply $\psi^2$ (rather then $|\psi|^2\equiv \psi
\psi^*$)  since the eigenfunctions are real.

\begin{figure}
\begin{center}
\includegraphics [width=0.9\columnwidth]{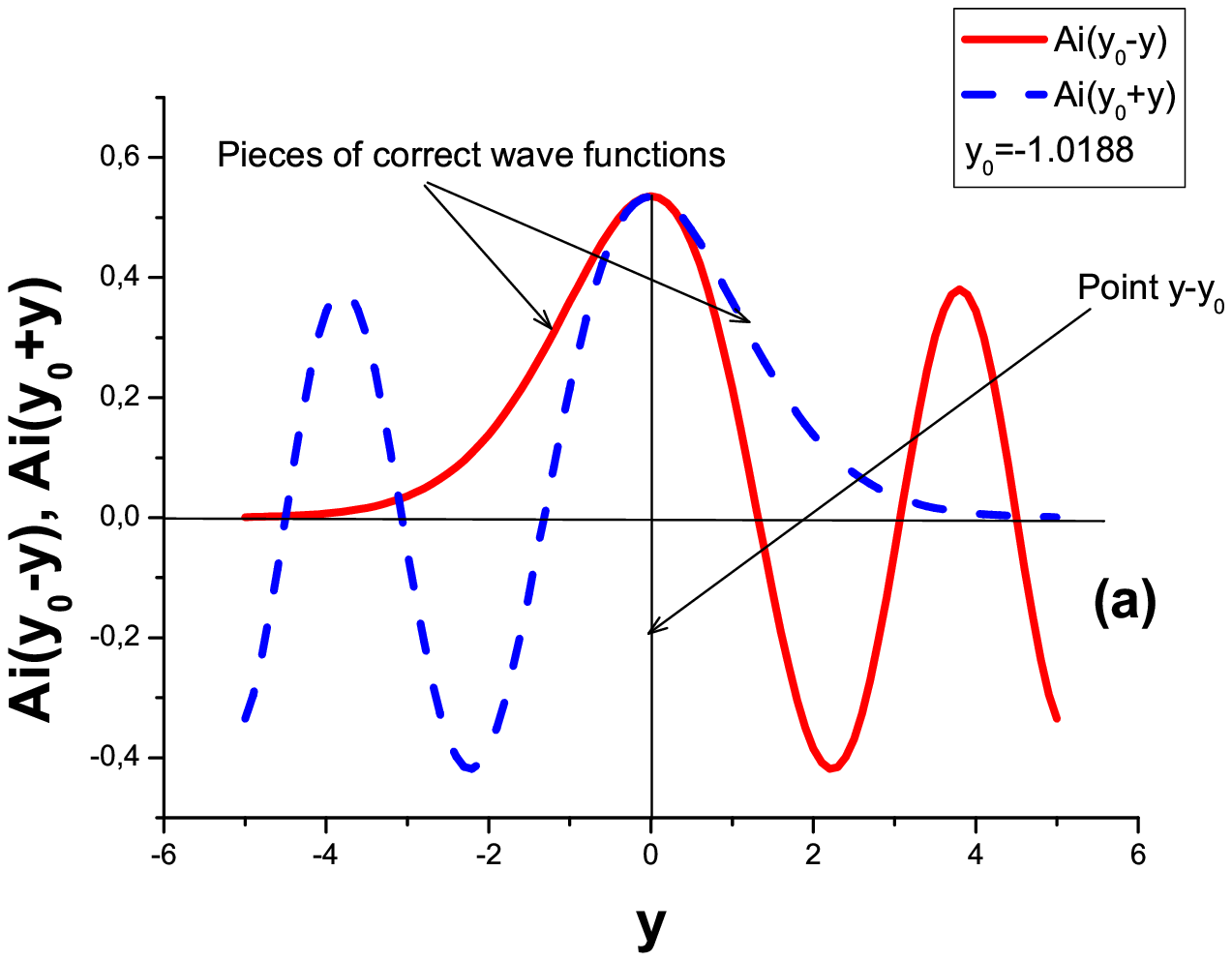}
\includegraphics [width=0.9\columnwidth]{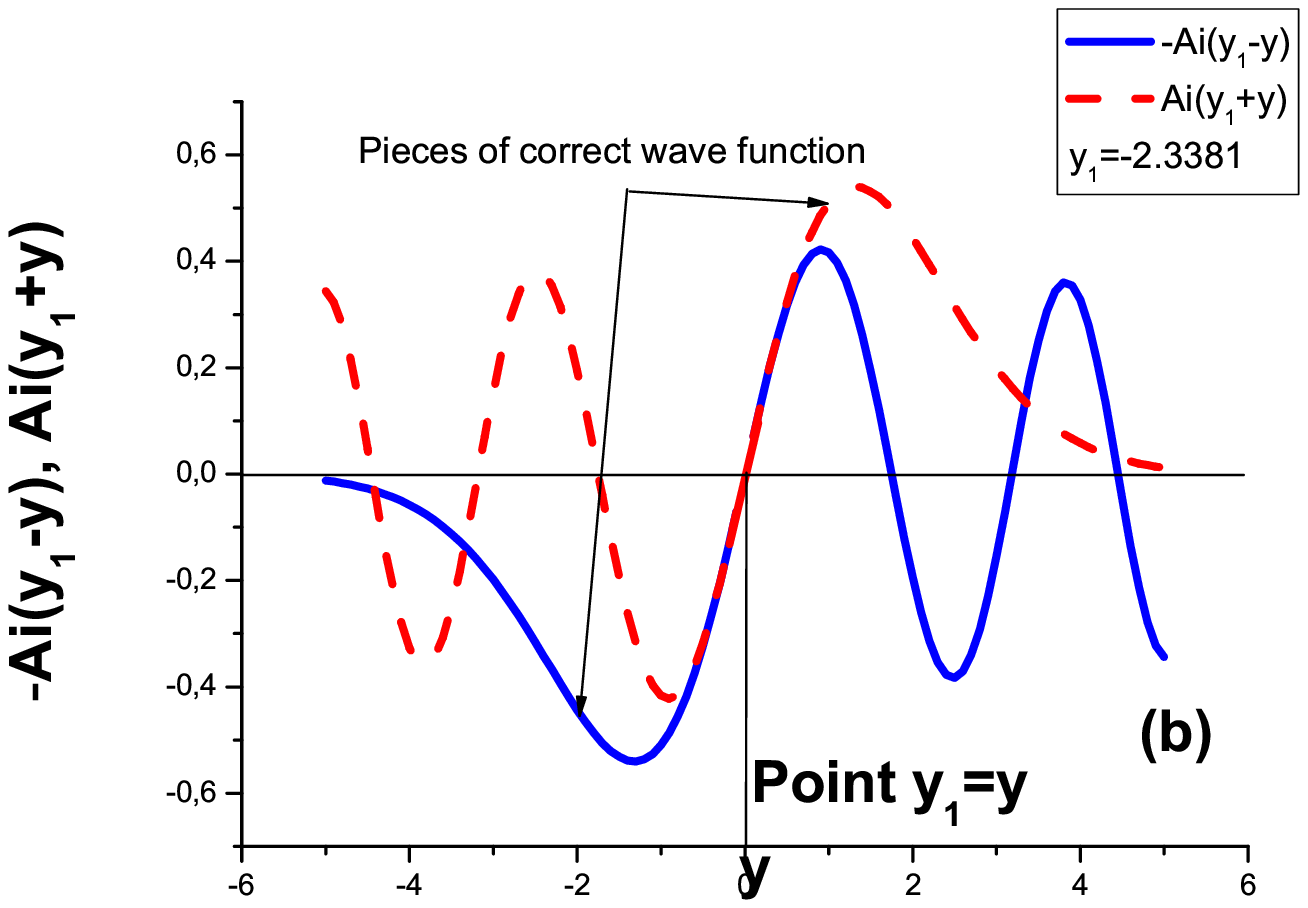}
\end{center}
\caption{The idea of the  wave functions construction; a- ground
state (and even $n$), b - first excited state (and odd $n$)}
\label{fig:ai}
\end{figure}

The method of construction of the above wave functions from initial Airy function is shown graphically in
Fig.\ref{fig:ai}. We link either function (odd $n$ - Fig.\ref{fig:ai}b) or its derivative (even $n$ - Fig.\ref{fig:ai}a) in one of the zeros. In other words, the wave function of $n$-th state is created by shifting the Airy function on
the positive half-axis to the right so that its n-th zero $y_n$ coincides with the origin  (zero) of the coordinate system.
 After this step we continue a  function to the negative half-axis either evenly (for even $n$) or oddly for odd
$n$.
The expression for $A_n$ is as follows
\begin{equation}\label{spe3}
A_n^2\left[\int_{-\infty}^0{\rm
{Ai}}^2(-y_n-y)dy+\int_0^{\infty}{\rm {Ai}}^2(-y_n+y)dy\right]=1.
\end{equation}
The evaluation of integrals \eqref{spe3} yields after some algebra
\begin{equation}\label{spe6}
 A_n=\frac{1}{\sqrt{2I_2}}\equiv
\frac{1}{\sqrt{2}}\left[y_n{\rm {Ai}}^2(-y_n)+{\rm
{Ai}}^{'2}(-y_n)\right]^{-1/2},
\end{equation}
where prime means derivative with respect to an argument.
Further simplifications of Eq. \eqref{spe6} are possible if we observe that for even $n$
${\rm {Ai}}^{'2}(-y_n)=0$ and for odd $n$ ${\rm {Ai}}^{2}(-y_n)=0$ so that
\begin{equation}
A_{n}=\left\{
\begin{array}{c}
\left[ \mathrm{Ai}^{\prime }(-y_{n})\sqrt{2}\right] ^{-1},\ n\
\text{is\
odd} \\
\left[ \mathrm{Ai}(-y_n)\sqrt{2y_n}\right] ^{-1},\ n\ \text{is\
even}.
\end{array}
\right.   \label{spe7}
\end{equation}

Also, the corresponding (dimensional) energy eigenvalues
$E_n\equiv {\cal V}_{0n}$ can be found from a condition that in the  $p$ space all
$p_n$ corresponding to zeros $y_n$ must be zero. In other words,
$x_n=y_n\zeta+\sigma\equiv 0$ or
\begin{eqnarray}
E_n=-\gamma \zeta y_n\equiv
|y_n|\left(\frac{\kappa }{2\gamma}\right)^{1/3}.\label{spe8}
\end{eqnarray}
Here we reflect the fact that zeros of the Airy function and its derivative are negative.

\begin{figure}
\begin{center}
\includegraphics [width=0.9\columnwidth]{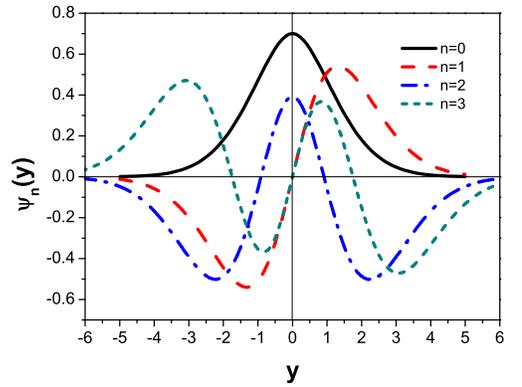}
\end{center}
\caption{Several  wave functions for the  linear modulus potential,
Eq. \eqref{rai2}.}\label{fig:sev}
\end{figure}

Now we are in a position to write the explicit form of several first wave functions of Eq. \eqref{rai2}. The ground state function is defined by the Eq. \eqref{frd17}. The first excited state ($n=1$) has the form
\begin{eqnarray}\label{n1}
&&    \psi _1(y)=A_1\left\{
\begin{array}{c}
{\rm Ai}(-y_1+y),\ y>0 \\
-{\rm {Ai}}(-y_1-y),\ y<0,
\end{array}
\right.\\
 &&A_1=\left[ \mathrm{Ai}'(-y_1)\sqrt{2}\right] ^{-1}, \ y_1 \approx
 2.3381.\nonumber
\end{eqnarray}
For the second excited state ($n=2$) we get
 \begin{eqnarray}\label{n2}
&&    \psi _2(y)=A_2\left\{
\begin{array}{c}
{\rm Ai}(-y_2+y),\ y>0 \\
{\rm {Ai}}(-y_2-y),\ y<0,
\end{array}
\right.\\
 &&A_2=\left[ \mathrm{Ai}(-y_2)\sqrt{2y_2}\right] ^{-1}, \ y_2 \approx
 3.2482.\nonumber
\end{eqnarray}
The  third excited state ($n=3$) reads
\begin{eqnarray}\label{n3}
&&    \psi _3(y)=A_3\left\{
\begin{array}{c}
{\rm Ai}(-y_3+y),\ y>0 \\
-{\rm {Ai}}(-y_3-y),\ y<0,
\end{array}
\right.\\
 &&A_1=\left[ \mathrm{Ai}'(-y_3)\sqrt{2}\right] ^{-1}, \ y_3 \approx
 3.2482.\nonumber
\end{eqnarray}
The functions \eqref{frd17}, \eqref{n1}  - \eqref{n3} are plotted in Fig. \ref{fig:sev}. The conformance with oscillation theorem (the $n$-th wave function of a discreet spectrum can have only $n$ zeros on its domain) \cite{landau} is seen.

\section{The approximate expressions for zeroes of Airy function and its derivative}

For large negative $y$ the function Ai$(y)$ has  the  following
asymptotic expansion
\begin{widetext}
\begin{eqnarray}\label{ai1}
&&{\rm Ai}(x\to-\infty)\approx
-\cos\left[\frac{\pi }{4}+\frac{2
x\sqrt{-x} }{3}\right] \frac{(-1)^{3/4}
\left(\frac{1}{x}\right)^{1/4}}{\sqrt{\pi }}-\frac{5 (-1)^{1/4}
\left(\frac{1}{x}\right)^{7/4}}{48 \sqrt{\pi }} \sin\left[\frac{\pi
}{4}+\frac{2 x\sqrt{-x}}{3}\right].
\end{eqnarray}
\end{widetext}
Equating series \eqref{ai1} to zero, we obtain the desired
analytical expression for zeros of the  Ai function. We observe that the
coefficient before $\sin$, proportional to $x^{-7/4}$,  decays at
infinity much faster then that preceding the  cosine. So, we simply
 equate to zero the argument of cosine function, getting
\begin{equation}
y_n= - \left(\frac{3\pi}{2}
\right)^{2/3} \left( {n + \frac{3}{4}} \right)^{2/3}.\label{ai2}
\end{equation}
It turns out that \eqref{ai2} gives a fairly good approximation for
zeros that begin at the  lowest eigenvalue $n=1$.
We compare exact and approximate \eqref{ai2} solutions:
\begin{eqnarray}
&&n=0 \quad y_0^{\rm exact}=-2.3381, \quad y_0^{\rm
appr}=-2.32025,\nonumber \\
&&n=1 \quad y_1^{\rm exact}=-4.0879, \quad y_1^{\rm
appr}=-4.08181,\nonumber \\
&&n=2 \quad y_2^{\rm exact}=-5.5206, \quad y_2^{\rm
appr}=-5.51716,\nonumber \\
&&...............................................\nonumber \\
&&n=8 \quad y_8^{\rm exact}=-11.0085, \quad y_8^{\rm
appr}=-11.0077,\nonumber \\
&&n=9 \quad y_9^{\rm exact}=-11.936, \quad y_9^{\rm
appr}=-11.9353. \label{ai3}
\end{eqnarray}

The same asymptotic analysis  can be  performed for the derivative of Airy function.
We end up with
\begin{equation}\label{ai4}
y_n^{(+)}= - \left(\frac{3\pi}{2} \right)^{2/3} \left( {n +
\frac{1}{4}} \right)^{2/3}.
\end{equation}
A comparison of exact and approximate roots goes as follows
\begin{eqnarray}
&&n=0 \quad y_0^{(+)\rm exact}=-1.0188, \quad y_0^{(+)\rm
appr}=-1.11546,\nonumber \\
&&n=1 \quad y_1^{(+)\rm exact}=-3.2482, \quad y_1^{(+)\rm
appr}=-3.26163,\nonumber \\
&&n=2 \quad y_2^{(+)\rm exact}=-4.8201, \quad y_2^{(+)\rm
appr}=-4.82632,\nonumber \\
&&...............................................\nonumber \\
&&n=8 \  \ y_8^{(+)\rm exact}=-11.4751, \ y_8^{(+)\rm
appr}=-11.4762,\nonumber \\
&&n=9 \  \  y_9^{(+)\rm exact}=-12.3848, \ y_9^{(+)\rm
appr}=-12.3857. \nonumber \\ \label{ai5}
\end{eqnarray}

\section{Fourier images of the wave functions}

Since our departure point in  Section III D was the Schr\"{o}dinger-type  eigenvalue problem in
momentum space, it turns out to be useful to discuss Fourier images
of the above eigenfunctions. We recall that the inverse (from momentum to coordinate space) Fourier transform is
defined as $f(x)=\frac{1}{2\pi}\int_{-\infty}^\infty f(p)e^{-ipx}dp$ .

For the ground state we  proceed accordingly. We substitute Eq. \eqref{frd17} into the Fourier integral to obtain

\begin{eqnarray}
&&\psi_0(x)=\frac{A_0}{2\pi}(I_1+I_2), \label{aif2}\\
&&I_1= \int_{-\infty}^0{\rm
Ai}(-y_0-p)e^{ipx}dp=\int_{-y_0}^\infty{\rm Ai}(t)e^{-ix(t+y_0)}dt,\nonumber \\
&&I_2= \int_0^{\infty}{\rm
Ai}(-y_0+p)e^{ipx}dp=\int_{-y_0}^\infty{\rm Ai}(t)e^{-ix(t+y_0)}dt, \nonumber \\
&&I_1+I_2=\int_{-y_0}^\infty{\rm
Ai}(t)\left[e^{-ix(t+y_0)}+e^{ix(t+y_0)}\right]dt. \nonumber
\end{eqnarray}

Hence, the  Fourier image of the  ground state wave function is determined by the above equation \eqref{aif6}.
For the higher even $n$ the above method yields
\begin{equation}\label{aif7}
\psi_{\rm even}(x)=\frac{A_n}{\pi}\int_{-y_n}^\infty{\rm Ai}(t)\cos x(t+y_n)dt,
\end{equation}
For odd states
\begin{eqnarray}
&&\psi_{\rm odd}(p)=\frac{A_n}{2\pi}(I_2-I_1),\label{aif8}\\
&&I_1\equiv \int_{-\infty}^0{\rm
Ai}(-y_n-p)e^{ipx}dp=
\int_{-y_n}^\infty{\rm Ai}(t)e^{-ip(t+y_n)}dt,\nonumber \\
&&I_2\equiv \int_0^{\infty}{\rm
Ai}(-y_n+p)e^{ipx}dp=\int_{-y_n}^\infty{\rm Ai}(t)e^{ix(t+y_n)}dt, \nonumber \\
&&I_2-I_1=A_n\int_{-y_n}^\infty{\rm
Ai}(t)\left[e^{ip(t+y_n)}-e^{-ip(t+y_n)}\right]dt.\nonumber
\end{eqnarray}
In other words, the Fourier images of odd  wave functions  have the form
\begin{equation}\label{aif12}
\psi_{\rm odd}(p)=i\frac{A_n}{\pi}\int_{-y_n}^\infty{\rm Ai}(t)\sin
p(t+y_0)dt.
\end{equation}
Here, $A_n$ are determined by Eqs. \eqref{spe7}. It is seen that
odd Fourier images are  imaginary odd functions. This
(imaginary coefficient) does not affect the  physical meaning of  $|\psi_{\rm odd}(p)|^2$ which
 is  a   probability density.


\begin{thebibliography}{9}
\bibitem{lev1} P. L\'{e}vy, {\it Processus stochastiques et mouvement Brownien}, Gauthier–Villars, Paris, 1965.
\bibitem{lev2} P. L\'{e}vy, {\it Th\'{e}orie de l'addition des variables al\'{e}atoires},
Gauthier-Villars, Paris, 1954.
\bibitem{risken} H. Risken, {\it The Fokker-Planck equation}, Springer-Verlag, Berlin, 1989,
\bibitem{stef} P. Garbaczewski and V. Stephanovich, Phys. Rev. E {\bf 80}., 031113, (2009)
\bibitem{sokolov} D. Brockmann and I. Sokolov, Chem. Phys. {\bf 284}, 409, (2002)
\bibitem{geisel} D. Brockmann and T. Geisel, Phys. Rev. Lett. {\bf 90}, 170601, (2003)
\bibitem{klafter} I. Eliazar and J. Klafter, J. Stat. Phys. {\bf 111}, 739, (2003)
\bibitem{mackey} A. Lasota and M.  C. Mackey, {\it Fractals and
noise: Stochastic aspects of dynamics}, Springer-Verlag, Berlin,
1995
\bibitem{olk0} P. Garbaczewski and R. Olkiewicz, J. Math. Phys. 37, 732, (1996)
\bibitem{olk} P. Garbaczewski and R. Olkiewicz,  J. Math. Phys. {\bf 40}, 1057, (1999)
\bibitem{olk1} P. Garbaczewski and R. Olkiewicz, J. Math. Phys. {\bf 41}, 6843, (2000)
\bibitem{klauder} P. Garbaczewski, J. R. Klauder and R. Olkiewicz, Phys. Rev. {\bf E 51}, 4114, (1995)
\bibitem{applebaum}  D. Applebaum, {\it L\'{e}vy processes and stochastic calculus}. Cambridge University Press, 2004
\bibitem{cufaro} N. Cufaro Petroni and M. Pusterla, Physica {\bf A 388}, 824, (2009)
\bibitem{laskin} N. Laskin, Phys. Rev. E {\bf 62}, 3135, (2000)
\bibitem{laskin1} N. Laskin, Phys. Rev. E {\bf 66}, 056108, (2002)
\bibitem{fogedby} S. Jespersen, R. Metzler and H. C. Fogedby, Phys. Rev. {\bf E 59}, 2736, (1999)
\bibitem{dubkov} A. A. Dubkov, B. Spagnolo and V. V. Uchaikin, Int. J. Bifurcations and Chaos  {\bf 18}, 2649, (2008)
\bibitem{chechkin} A. Chechkin et al, Chem. Phys.  {\bf 284}, 233, (2002)
\bibitem{chechkin1} A. Chechkin et al, J. Stat. Phys. {\bf 115}, 1505, (2004)
\bibitem{ux} To solve the corresponding equations numerically, we use simple Euler scheme for time derivatives
along with (at each time step) numerical calculation of Cauchy principal value of integrals for evaluation of fractional derivative $|\nabla|$.
\bibitem{robinett}  R. W. Robinett, Am. J. Phys. {\bf 63}, 823, (1995)
\bibitem{ruiz} J. S\'{a}nchez-Ruiz, Phys. Lett. {\bf A 226}, 7, (1997)
\bibitem{moreno} P. S\'{a}nchez-Moreno, R. J. Y\'{a}$\tilde{n}$ez and V. Buyarov, J. Phys. A: Math. Gen. {\bf 38}, 9969, (2005)
\bibitem{landau} L. D. Landau and E. M. Lifshitz, {\it  Quantum mechanics}, Addison-Wesley, NY, 1965


\end{thebibliography}
\end{document}